\documentclass[12pt]{article}
\begin{document}
\begin{center}
{\large ENERGY CRISIS OR A NEW SOLITON \\
IN THE NONCOMMUTATIVE $CP(1)$ MODEL? }\\
\vskip 2cm
Subir Ghosh\\
\vskip 1cm
Physics and Applied Mathematics Unit,\\
Indian Statistical Institute,\\
203 B. T. Road, Calcutta 700108, \\
India.
\end{center}
\vskip 3cm
{\bf Abstract:}\\
The Non-Commutative (NC) $CP(1)$ model is studied from field theory perspective. Our
formalism and definition of the NC $CP(1)$ model differs crucially from the existing one
\cite{nccp}.

 Due to the $U(1)$ gauge invariance, the Seiberg-Witten map is used to convert the NC action
 to an action in terms of ordinary spacetime degrees of freedom and the subsequent theory is
 studied. The NC effects appear as (NC parameter)
 $\theta $-dependent interaction terms. The expressions for static energy, obtained from both
 the symmetric and canonical forms of the energy momentum tensor, are {\it{identical}}, when
 only spatial noncommutativity is present. Bogomolny analysis reveals a lower bound in the
 energy in an unambiguous way, suggesting the presence of a new soliton. However, the BPS
 equations saturating the bound are not compatible to the full variational equation of motion.
 This indicates that the definitions of the energy momentum tensor for this particular  NC
 theory, (the NC theory is otherwise consistent and well defined), are inadequate, thus leading to the "energy crisis".

 A collective coordinate analysis corroborates the above observations. It also
 shows that the above mentioned mismatch between the BPS equations and the variational
 equation of motion is small.

\newpage

\begin{center}
{\bf{Introduction}}
\end{center}

Non-Commutative (NC) field theories have turned into a hotbed of
research activity after its connection to low energy string
physics was elucidated by Seiberg and Witten \cite{sw,rev}.
Specifically, the open string boundaries, attached to D-branes
\cite{pol}, in the presence of a two-form background field, turn
into NC spacetime coordinates \cite{sw}. (This phenomenon has been
recovered from various alternative viewpoints \cite{others}.) The
noncommutativity induces an NC D-brane world volume and hence
field theories on the brane become NC field theories.

Studies in NC   field theories have revealed unexpected features,
such as UV-IR mixing  \cite{min1}, soliton solutions in higher
dimensional scalar theories \cite{min2}, to name a few. The inherent non-locality, (or equivalently the introduction of a length scale by $\theta$ - the noncommutativity parameter), of the NC field theory is manifested through these peculiar properties, which  are absent in the corresponding ordinary spacetime
theories. Also, solitons in NC $CP(1)$ model have been found
\cite{nccp}, very much in analogy to their counterpart in ordinary
spacetime. The present work also deals with the search for the solitons in the  NC $CP(1)$
model. The difference between our  field theoretic analysis and
the existing framework \cite{nccp} is explained below. In fact, we
closely follow  the conventional field theoretic approach in
ordinary spacetime \cite{vij}. The Seiberg-Witten map \cite{sw}
plays a pivotal role in our scheme. The Bogolmolny analysis of
the static energy reveals a lower bound, protected by topological
considerations. However, we encounter a small discrepancy between
the BPS equations and the variational equation of motion. Although "small" in
an absolute sense, the mismatch is conceptually significant for the reasons elaborated below.
The above conclusions, drawn from field theoretic analysis,
will be corroborated and quantified explicitly  in a collective
coordinate framework.

 It appears natural to attribute
the above mentioned problem to the definition of the energy functional of the NC $CP(1)$
model in particular, and of the NC field theories in general, (since the BPS equations are
derived directly from the energy of the system).
 As it is well known, there are complications in the
definition of the Energy-Momentum  (EM) tensor in NC field theory
\cite{grim,das}. In general, it is not possible to obtain a
symmetric, gauge invariant and conserved EM tensor. There are two
forms of EM tensor in vogue: a manifestly symmetric form
\cite{das}, obtained from the variation of the action with respect
to the metric, and the canonical form \cite{grim}, following the
Noether prescription. The former is covariantly conserved whereas
the latter is conserved. Interestingly, we find that in the
particular case that we are considering, that is NC $CP(1)$ model
in 2+1-dimensions, with only spatial noncommutativity,  {\it{expressions for the
static energy, obtained  from both the derivations, are
identical}}. Moreover the expression for energy is gauge
invariant. We show that there is a Bogomolny like lower bound in
the energy. However, the subsequently derived BPS equations (that
saturate the lower bound), does not fully satisfy the  equation of
motion.

Let us put our work in its proper perspective. As such, our result
in no way questions the consistency of the existing literature
\cite{nccp} on $CP(1)$ solitons since our model differs crucially
from the one considered in \cite{nccp}. In particular, we have
adopted a different NC generalization of the $CP(1)$ constraint.
(We have provided conceptual and technical reasons for allowing
such a difference.) Thus it is not expected that the results of
\cite{nccp} will be reproduced. In fact, the energy profile of the
localized structure that we uncover, is more sharply peaked and
has an $O(\theta )$ correction, with respect to the ordinary
spacetime $CP(1)$ soliton. (This will be made explicit in the
collective coordinate analysis at the end.) Surprisingly, the BPS
equations remain unchanged, although the variational equation of motion is altered.

All the same, we emphasize that, we have provided a well defined
NC gauge theory, which conforms to the expected features of such a
system. Hence, the clash between the BPS equations and the
equation of motion that is revealed here, can have a deeper
bearing on the structure of a general NC gauge theory, indicating
that the traditional lore of field theory in ordinary spacetme
should be applied with greater care in the context of NC field
theory.

Our methodology and its difference from the existing one \cite{nccp} is explained below.
There are two basic approaches in studying an
NC field theory: \\
(i) The appropriate NC field theory is constructed in terms of NC
analogue fields $(\hat \psi )$ of the fields $(\psi)$ with the
replacement of ordinary products of fields $(\psi \varphi )$, by
the Moyal-Weyl $*$-product $(\hat \psi *\hat \varphi )$,
\begin{equation}
\hat \psi(x)*\hat
\varphi(x)=e^{\frac{i}{2}\theta_{\mu\nu}\partial_{\sigma_{\mu}}\partial_{\xi_{\nu}}}\hat
\psi(x+\sigma )\hat \varphi(x+\xi )\mid_{\sigma =\xi =0} =\hat
\psi(x)\hat \varphi(x)+
\frac{i}{2}\theta^{\rho\sigma}\partial_{\rho}\hat
\psi(x)\partial_{\sigma}\hat \varphi(x)+~O(\theta^{2}). \label{mw}
\end{equation}
The {\it{ hatted}} variables are NC degrees of freedom. We take
$\theta^{\rho\sigma}$   to be a real constant antisymmetric
tensor, as is customary \cite{sw}, (but this need not always be
the case \cite{sny}). The NC spacetime follows from the above
definition,
\begin{equation}
[x^{\rho},x^{\sigma}]_{*}=i\theta^{\rho\sigma}.
\label{nc}
\end{equation}
Note that the effects of spacetime noncommutativity has been
accounted for by the  introduction of the $*$-product.
For gauge theories the Seiberg-Witten Map \cite{sw} plays a crucial role in  connecting $\hat \phi (x) $ to $\phi (x)$. This formalism allows us to study the effects of noncommutativity as $\theta^{\rho\sigma}$ dependent interaction terms in an ordinary spacetime field theory format. This is the prescription we will follow.\\
(ii) An alternative framework is to treat the NC theories as
systems of operator  valued fields and to directly work with
operators on the quantum phase space, characterized by the
noncommutativity condition (\ref{nc}). On the NC plane, the
coordinates satisfy a Heisenberg algebra
$[x^1,x^2]_{*}=x^{1}*x^{2}-x^{2}*x^{1}=i\theta
^{12}=i\epsilon^{12}\theta =i\theta $ which in the complex
coordinates reduces to the creation annihilation operator algebra
for the simple Harmonic Oscillator. Thus to a function in the NC
spacetime, through Weyl transform, one associates an operator
acting on the Hilbert space, in a basis of a simple Harmonic
Oscillator eigenstates.

The investigations on the NC $CP^{1}$ solitons carried out so far
\cite{nccp,muru}  exploit the latter method. As it turns out, a
major advantage is that structurally, the NC system with its
dynamical equations, energy functionals etc., are similar to their
ordinary spacetime counterpart. This happens because the
$*$-products of (i) are replaced by operator products in (ii) and
the spacetime integrals are replaced by trace over the basis
states in the hilbert space.

In the present work, our aim is to study the NC $CP^{1}$ solitons
in the former  field theoretic approach. From past experiences
\cite{sg} we know this to be a perfectly viable formalism. Indeed,
since the NC spacetime physics is not that much familiar or well
understood, it is imperative that one explores different avenues
to reach the same goal, to gain further insights. Also we would
like to point out that since solitons are already present in the
$CP^{1}$ model at $\theta =0$ (i.e. ordinary spacetime), unlike
the noncommutative solitons of the scalar theory \cite {min2} it
is natural to analyze  the fate  of the solitons under a small
perturbation, (which is a small value of $\theta$ in the present
case). The small $\theta$-results of \cite{nccp,muru} are
perfectly well defined.

The paper is organized as follows: Section {\bf {II}} contains a
short recapitulation of the $CP(1)$ solitons. This will help us
fix the notations and in fact, identical procedure will be pursued
in the NC theory as well. The detailed construction of our version
of the NC $CP(1)$ model is provided in section {\bf{III}}. Section
{\bf {IV}} discusses the energy momentum tensor of the model.
Section {\bf{V}} consists of the Bogomolny analysis in the NC
theory. Section {\bf{VI}} is devoted to the collective coordinate
analysis. The paper ends with a conclusion in Section {\bf{VII}}.
\vskip .5cm
\begin{center}
{\bf{Section II - $CP(1)$ Soliton: a brief digression}}
\end{center}

Let us digress briefly on the BPS solitons of $CP(1)$ model in
ordinary spacetime.  Later we will proceed with the NC theory in an identical fashion.
The gauge invariant action,
\begin{equation}
S=\int d^{3}x~[(D^{\mu}\phi )^{\dag}D_{\mu}\phi +\lambda (\phi^{\dag}\phi -1)],
\label{ac}
\end{equation}
where $D_{\mu}\phi =(\partial_{\mu}-iA_{\mu})\phi$ defines the
covariant derivative  and the multiplier $\lambda$ enforces the
constraint, the equation of motion for $A_{\mu}$ leads to the
identification,
\begin{equation}
A_{\mu}=-i\phi^{\dag}\partial_{\mu}\phi .
\label{nca1}
\end{equation}
Since the "gauge field" $A_{\mu}$ does not have any independent
dynamics one is allowed  to make the above replacement directly in
the action. Obviously the  infinitesimal gauge transformation of
the variables are,
\begin{equation}
\delta \phi^{\dag}=-i\lambda \phi^{\dag} ~;~~\delta \phi=i\lambda
\phi~; ~~\delta A_{\mu}=\partial_{\mu}\lambda . \label{gt}
\end{equation}
From the EM tensor
\begin{equation}
T_{\mu\nu}=(D_{\mu}\phi)^{\dag}D_{\nu}\phi
+(D_{\nu}\phi)^{\dag}D_{\mu} \phi
-g_{\mu\nu}(D^{\sigma}\phi)^{\dag}D_{\sigma}\phi , \label{t1}
\end{equation}
the total energy can be expressed in the form,
\begin{equation}
E=\int d^{2}x~(\mid D_{0}\phi \mid^{2}+\mid (D_{1}\pm iD_{2})\phi \mid^{2})\pm2\pi N ,
\label{e}
\end{equation}
where the last term denotes the topological charge
\begin{equation}
N\equiv \int d^{2}x~n(x)=\frac{1}{2\pi i}\int d^{2}x~\epsilon^{ij} (D_{i}\phi )^{\dag} D_{j}\phi =\int d^{2}x~\frac{1}{4\pi }\epsilon^{ij}F_{ij}=\int d^{2}x~\frac{1}{2\pi }F_{12},
\label{n}
\end{equation}
corresponding to the conserved topological current. The Bogomolny
bound follows from  (\ref{e}),
\begin{equation}
E\geq2\pi \mid N\mid ,
\label{bog}
\end{equation}
with the following saturation conditions (BPS equations) obeyed by the soliton,
\begin{equation}
\mid D_{0}\phi \mid^{2}=\mid (D_{1}\pm iD_{2})\phi \mid^{2}=0 .
\label{bog1}
\end{equation}
It can be checked that the solutions of the BPS equations belong to a subset of the full set of solutions, that satisfy the variational equation of motion.
\vskip .5cm
\begin{center}
{\bf{Section III - Construction of the NC $CP(1)$ model}}
\end{center}

Let us now enter the noncommutative spacetime. The first task is
to generalize the scalar  gauge theory (\ref{ac}) to its NC
version, keeping in mind that the latter must be $*$-gauge
invariant. The NC action is,
\begin{equation}
\hat S=\int d^{3}x~(\hat D^{\mu}\hat\phi )^{\dag}*\hat
D_{\mu}\hat\phi =\int d^{3}x~ (\hat D^{\mu}\hat\phi )^{\dag}\hat
D_{\mu}\hat\phi , \label{ncac}
\end{equation}
where the NC covariant derivative is defined as
$$\hat D_{\mu}\hat\phi =\partial_{\mu}\hat\phi-i\hat A_\mu*\hat\phi .$$
Depending on the positioning
of $\hat A_{\mu}$ and $\hat \phi$, the covariant derivative can act in three ways ,
\begin{equation}
\hat D_{\mu}\hat\phi
\begin{array}[t]{l}
 =\partial_{\mu}\hat\phi-i\hat A_\mu*\hat\phi \\
=\partial_{\mu}\hat\phi+i\hat\phi*\hat A_\mu \\
=\partial_{\mu}\hat\phi-i(\hat\phi*\hat A_\mu -\hat A_\mu*\hat\phi )
\end{array}
\label{ncov}
\end{equation}
which are termed respectively as fundamental, anti-fundamental and
adjoint representations. We have chosen the fundamental one.
{\footnote{This is the first difference  between our model and
\cite{nccp} who use the anti-fundamental representation. In fact,
in \cite{nccp}, it is difficult to proceed with the fundamental
definition \cite{yang}. On the other hand, in the present work,
the choice between the first and second definition is not very
important as it affects the overall sign of $\theta$ only. Similar
type of situation prevails in \cite{gs,sg}.}} Notice that for the
time being we have not considered the target space (CP(1))
constraint. We will return to this important point later. The NC
action (\ref{ncac}) is invariant under the $*$-gauge
transformations,
\begin{equation}
\hat \delta \hat\phi^{\dag}=-i\hat\lambda *\hat\phi^{\dag}
~;~~\hat\delta  \hat\phi=i\hat\lambda *\hat\phi~;~~\hat\delta \hat
A_{\mu}=\partial_{\mu}\hat\lambda +i[\hat \lambda ,\hat
A_{\mu}]_{*}. \label{ncgt}
\end{equation}
We now exploit the Seiberg-Witten Map \cite{sw,jur} to revert back
to the ordinary  spacetime degrees of freedom. The explicit
identifications between NC and ordinary spacetime counterparts of
the fields, to the lowest non-trivial order in $\theta$ are,
$$
\hat A_{\mu}=A_{\mu}+\theta^{\sigma\rho}A_{\rho}(\partial_{\sigma}
A_{\mu}-\frac{1}{2} \partial_{\mu} A_{\sigma})
$$
\begin{equation}
\hat \phi =\phi -\frac{1}{2}\theta^{\rho\sigma} A_{\rho}\partial_{\sigma}\phi ~;~~
\hat \lambda = \lambda -\frac{1}{2}\theta^{\rho\sigma} A_{\rho}\partial_{\sigma} \lambda .
\label{swm}
\end{equation}
As stated before, the "`hatted"' variables on the left are NC
degrees of  freedom and gauge transformation parameter. The higher
order terms in $\theta$ are kept out of contention as there are
certain non-uniqueness involved in the $O(\theta^{2})$ mapping.
The significance of the Seiberg-Witten map is that under an NC or
$*$-gauge transformation of $\hat A_{\mu}$ by,
$$\hat\delta \hat A_{\mu}=\partial_{\mu}\hat \lambda +i[\hat\lambda ,\hat A_{\mu}]_{*},$$
$A_{\mu}$ will undergo the transformation $$\delta A_\mu
=\partial_{\mu}\lambda . $$ Subsequently, under this mapping, a
gauge invariant object in conventional spacetime will be mapped to
its NC counterpart, which will be $*$-gauge invariant. This is
crucial as it ensures that the ordinary spacetime action that  we
recover from the NC action (\ref{ncac}) by applying the
Seiberg-Witen Map will be gauge invariant. Thus the NC action
(\ref{ncac}) in ordinary spacetime variables reads,
\begin{equation}
\hat S =\int d^{3}x [( D^{\mu}\phi )^{\dag} D_{\mu}\phi +
\frac{1}{2}\theta^{\alpha\beta} \{F_{\alpha\mu}((D_{\beta}\phi
)^{\dag} D^{\mu}\phi + (D^{\mu}\phi )^{\dag} D_{\beta}\phi
)-\frac{1}{2}F_{\alpha\beta}(D^{\mu}\phi )^{\dag} D_{\mu}\phi \}]
\label{ncac1}
\end{equation}
The above action is manifestly gauge invariant. The equation of motion now satisfied by $A_{\mu}$
is,
$$
i(-2iA_{\mu}\phi^{\dag}\phi +\phi^{\dag}\partial_{\mu}\phi
-\partial_{\mu}\phi^{\dag}\phi )
(1-\frac{1}{2}\theta^{\alpha\beta}F_{\alpha\beta})
$$
$$
+\frac{1}{2}\theta_{\alpha\mu}[\partial^{\alpha}\{(D^{\beta}\phi
)^{\dag}D_{\beta}\phi \}   -\partial_{\beta}\{(D^{\alpha}\phi
)^{\dag}D^{\beta}\phi +(D^{\beta}\phi )^{\dag}D^{\alpha}\phi \}]$$
\begin{equation}
-\frac{1}{2}\theta_{\alpha\beta}[\partial^{\alpha
}\{(D^{\beta}\phi )^{\dag}D_{\mu}\phi +(D_{\mu}\phi
)^{\dag}D^{\beta}\phi )+iF^{\alpha
}_{~\mu}(\phi^{\dag}D^{\beta}\phi -(D^{\beta}\phi)^{\dag}\phi
\}]=0 \label{ncA}
\end{equation}
Remember that so far we have not introduced the $CP^{1}$ target space constraint
in the NC spacetime setup. {\it{Let us assume the constraint to be identical to the
ordinary spacetime one,}} i.e., {\footnote{This is the second difference between our
model and that of \cite{nccp}, where $\theta$-correction terms are present in the $CP(1)$
constraint. This is a serious difference as it drastically alters the structures of the
model in \cite{nccp} from ours. Apart from the conceptual reasoning given above, there also
appears a technical compulsion. We would like to obtain perturbative $\theta$-corrections to the ordinary spacetime $CP(1)$ model. Incorprtating $\theta$-corrections in the $CP(1)$ constraint as in \cite{nccp} in our system will lead to a differential equation for the multiplier $\lambda$, instead of an algebraic one as in the ordinary spacetime case. This will change the $\phi$-equation of motion in a qualitative way. We stress that our model is a perfectly well defined NC theory which, incidentally, is distinct from the existing NC $CP(1)$ model \cite{nccp}. }}
\begin{equation}
\phi ^{\dag}\phi =1.
\label{cp}
\end{equation}
The reasoning is as follows. Primarily, after utilizing the
Seiberg-Witten Map,  we have returned to the ordinary spacetime
and its associated dynamical variables and the effects of
noncommutativity appears only as additional interaction terms in
the action. Hence it is natural to keep the $CP^{1}$ constraint
unchanged. Alternatively, the above assumption can also be
motivated in a roundabout way. Remember that the the $CP(1)$
constraint has to be introduced in a $*$-gauge invariant way. In
order to introduce the $CP^{1}$ constraint directly in the NC
action (\ref{ncac}) or (\ref{ncac1}),  the constraint term
$\int~\lambda (\phi ^{\dag}\phi -1)$ has to be generalized to a
$*$-gauge invariant one by the application of the (inverse)
Seiberg-Witten Map. This is quite straightforward but needless
because as soon as we apply the Seiberg-Witten Map to the
$*$-gauge invariant constraint term, we recover the earlier
ordinary spacetime constraint.

This allows us to write,
\begin{equation}
A_{\mu}=-i\phi^{\dag}\partial_{\mu}\phi +a_{\mu}(\theta)
\label{nca}
\end{equation}
with $a_{\mu}$ denoting the $O(\theta )$ correction,  obtained
from (\ref{ncA},\ref{cp}). For $\theta =0$. $A_{\mu}$ reduces to its original form.
Note that $a_{\mu}$ is {\it {gauge invariant}}.
Thus the $U(1)$ gauge transformation of  $A_{\mu}$  remains
intact, at least to  $O(\theta)$.
 Keeping in mind the constraint $\phi^{\dag}\phi =1$,
 let us now substitute(\ref{nca}) in the NC action (\ref{ncac1}).
 Since we are concerned only with the $O(\theta)$ correction, in
 the $\theta$-term of the action, we can use $A_{\mu}=-i\phi^{\dag}\partial_{\mu}\phi$.
 However, in the first term in the action,    we must incorporate the full expression
 for $A_{\mu}$ given in (\ref{nca}). Remarkably, the constraint condition conspire to
 cancel the effect of the $O(\theta)$ correction term $a_{\mu}$. Finally it boils down
 to the following: the action for the  NC $CP^{1}$ model to $O(\theta )$ is given by
 (\ref{ncac1}) with the identification $A_{\mu}=-i\phi^{\dag}\partial_{\mu}\phi$ and
 $\phi^{\dag}\phi =1$.

\vskip .5cm
\begin{center}
{\bf{Section IV - Energy-momentum tensor for the NC $CP(1)$ model}}
\end{center}
 Our aim is to study the possibility of soliton solutions for the
action (\ref{ncac1}).   Let us try to derive the Bogolmony bound
and BPS equations in the present case. The first task is to compute the EM tensor.

We follow \cite{das} in computing the symmetric form of the
EM tensor  by coupling the model with a weak
gravitational field and get,
$$
T^{S}_{\mu\nu}=(1-\frac{1}{4}\theta^{\alpha\beta}F_{\alpha\beta})[(D_{\mu}\phi)
^{\dag}D_{\nu}\phi +(D_{\nu}\phi)^{\dag}D_{\mu}\phi
-g_{\mu\nu}(D^{\sigma}\phi)^{\dag}D_{\sigma}\phi
]+\frac{1}{2}\theta^{\alpha\beta}[F_{\alpha\mu}((D_{\beta}\phi)^{\dag}D_{\nu}\phi
+(D_{\nu}\phi)^{\dag}D_{\beta}\phi )$$
\begin{equation}
+ F_{\alpha\nu}((D_{\beta}\phi)^{\dag}D_{\mu}\phi
+(D_{\mu}\phi)^{\dag}D_{\beta}\phi )
-g_{\mu\nu}F_{\alpha\sigma}((D_{\beta}\phi)^{\dag}D^{\sigma}\phi
+(D^{\sigma}\phi) ^{\dag}D_{\beta}\phi )]. \label{t}
\end{equation}
The $T^{S}_{\mu\nu}$ stands for the symmetric form of the EM
tensor. In the static situation, $$\dot \phi =0~~\rightarrow
A_{0}=F_{0i}=D_{0}\phi =0$$ and the static energy density
simplifies to,
\begin{equation}
T^{S}_{00}=(1+\frac{1}{2}\theta^{12}F_{12})(D^i\phi)^{\dag}D^i\phi ,
\label{en}
\end{equation}
where $F_{\mu\nu}$ is also expressible in the form
$$F_{\mu\nu}=-i[(D_{\mu}\phi) ^{\dag}D_{\nu}\phi
-(D_{\nu}\phi)^{\dag}D_{\mu}\phi ]
=-i[(\partial_{\mu}\phi)^{\dag}\partial_{\nu}\phi
-(\partial_{\nu}\phi)^{\dag}\partial_{\mu}\phi ].$$

Now we discuss the canonical form of the EM tensor. Remembering
that the indices of adjacent $\phi$'s are summed, the expanded
form of the Lagrangian is,
$$
\hat L=(D^{\mu}\phi )^{\dag}D_{\mu}\phi +\lambda (\phi^{\dag}\phi
-1)-\frac{i}{2}\theta^
{\alpha\beta}[2\partial_{\mu}\phi^{\dag}\partial_{\beta}\phi
\partial_{\alpha}\phi^{\dag}\partial^{\mu}\phi
$$
\begin{equation}
 +2\partial_{\mu}\phi^{\dag}\partial_{\beta}\phi \phi^{\dag}\partial_{\alpha}\phi\phi^{\dag}\partial^{\mu}\phi -2\partial_{\beta}\phi^{\dag}\partial_{\mu}\phi \phi^{\dag}\partial_{\alpha}\phi\phi^{\dag}\partial^{\mu}\phi
-\partial_{\alpha}\phi^{\dag}\partial_{\beta}\phi
\partial^{\mu}\phi^{\dag}\partial_{\mu}\phi -
\partial_{\alpha}\phi^{\dag}\partial_{\beta}\phi \phi^{\dag}\partial^\mu \phi \phi^{\dag}\partial_{\mu}\phi ].
\label{lag}
\end{equation}
The canonical energy-momentum tensor is,
$$T_{\mu\nu}=\frac{\delta \hat L}{\delta (\partial^{\mu}\phi ^{\dag})}\partial_{\nu}\phi^{\dag}
+\frac{\delta \hat L}{\delta (\partial^{\mu}\phi )}\partial_{\nu}\phi  -g_{\mu\nu}\hat L $$
$$ =(1-\frac{1}{4}\theta^{\alpha\beta}F_{\alpha\beta} )
(D_{\mu}\phi^{\dag}D_{\nu}\phi +(\mu \leftrightarrow \nu ))$$
$$-i\theta^{\alpha\beta}[\partial_{\nu}\phi^{\dag}\partial_{\beta}\phi (\partial_{\alpha}\phi^{\dag}\partial_{\mu}\phi +\phi^{\dag}\partial_{\alpha}\phi \phi^{\dag}\partial_{\mu}\phi )
+(\mu \leftrightarrow \nu )
-\partial_{\beta}\phi^{\dag}\partial_{\nu}\phi \phi^{\dag} \partial_{\alpha}\phi \phi^{\dag}\partial_{\mu}\phi -(\mu \leftrightarrow \nu )]$$
$$-i\theta_{\mu\alpha}[(-\frac{1}{2}\partial_{\nu}\phi^{\dag}\partial^{\alpha}\phi +\frac{1}{2}\partial^{\alpha}\phi^{\dag}\partial_{\nu}\phi )(D^{\mu}\phi )^{\dag}D_{\mu}\phi
-\partial_{\sigma}\phi^{\dag}\partial_{\nu}\phi (\partial^{\alpha}\phi^{\dag}\partial^{\sigma}\phi +\phi^{\dag}\partial^{\alpha}\phi \phi^{\dag}\partial^{\sigma}\phi ) $$
\begin{equation}
+\partial_{\sigma}\phi^{\dag}\partial^{\alpha}\phi (\partial_{\nu}\phi^{\dag}\partial^{\sigma}\phi +\phi^{\dag}\partial_{\nu}\phi \phi^{\dag}\partial^{\sigma}\phi )
+\partial_{\nu}\phi^{\dag}\partial_{\sigma}\phi \phi^{\dag}\partial^{\alpha}\phi \phi^{\dag}\partial^{\sigma}\phi -
\partial^{\alpha}\phi^{\dag}\partial_{\sigma}\phi \phi^{\dag}\partial_{\nu}\phi \phi^{\dag}\partial^{\sigma}\phi ]-g_{\mu\nu}\hat L.
\label{cant}
\end{equation}
Note that $\theta^{\alpha\beta}$ part is symmetric. In the energy
density $T_{00}$  the contribution coming from the non-symmetric
parts in the $\theta$-contribution drop out if only space-space
noncommutativity is assumed, i.e. $\theta^{0i}=0$. For this
special case, in the static limit, the above $\theta$-contribution
completely drops out  and the energy density reduces to
\begin{equation}
T^{N}_{00}=-\hat L=(1+\frac{1}{2}\theta^{12}F_{12})(D^i\phi)^{\dag}D^i\phi.
\label{tn}
\end{equation}
Clearly this is identical to the static energy (\ref{en}) obtained
from the the  symmetric form $T^{S}_{\mu\nu}$. Indeed, it is
satisfying that in this particular case, both the canonical and
symmetric forms of the EM tensor lead to the same
expression of the static energy, which is manifestly gauge
invariant and conserved (as it comes from the canonical form).

 Interestingly to $O(\theta)$, the noncommutativity effect factors out from the ordinary
 spacetime result. Also notice that in the two spatial dimensions that we are considering,
 the $\theta$-term in the energy density is proportional to the topological charge density
 $n(x)$ in (\ref{n}). This is because the expression for the topological current remains
 unchanged since the dynamical variables as well as the $CP(1)$ constraint is unaltered
 in our model. We specialize to only space-space noncommutativity, $\theta ^{ij}=\theta
 \epsilon ^{ij}~,~\theta ^{0i}=0$, and  find,
\begin{equation}
T_{00}^{(\theta)}=\pi \theta n(x) (D^i\phi)^{\dag}D^i\phi ,
\label{tn1}
\end{equation}
where the superscript $S$ or $N$ is dropped.

\vskip .5cm
\begin{center}
{\bf{Section V - Analysis of the Bogomolny bound}}
\end{center}
In order to obtain
the Bogomolny bound, we follow the same procedure as that of the
$CP^{1}$ model in ordinary spacetime and rewrite the static energy
functional in the following form,
$$
E=\int d^{2}x~(1+\pi\theta n)[\mid (D_{1}\pm iD_{2})\phi \mid^{2}\pm2\pi n]$$
$$
=\int d^{2}x~[(1+\pi \theta n)\mid (D_{1}\pm iD_{2})\phi
\mid^{2}\pm2\pi n \pm 2\pi  ^2\theta n^{2}]$$
\begin{equation}
=\int d^{2}x~(1+\frac {1}{2}\pi \theta n(x))^2\mid (D_{1}\pm iD_{2})\phi \mid^{2} \pm2\pi N \pm 2\pi ^2\theta \int d^{2}x~ n^{2}(x) +O(\theta ^2).
\label{nvb}
\end{equation}
 Now individually all the terms in the energy expression are positive definite.
 Hence we obtain the Bogomolny bound to be
 \begin{equation}
 E\geq N+2\pi ^2\theta \int d^{2}x~ n^{2}(x)
 \label{bogol}
 \end{equation}
 and the saturation condition is
 \begin{equation}
 (1+\frac {1}{2}\pi \theta n(x))^2\mid (D_{1}\pm iD_{2})\phi \mid^{2}=0.
 \label{sat}
 \end{equation}
The BPS equation turns out to be,
\begin{equation}
D^1\phi =\pm iD^2\phi.
\label{bps}
 \end{equation}
Thus we find that the BPS equation remains unchanged and there is
a $O(\theta)$  correction in the static energy of the soliton.
Note that both of the  above results do not agree with
\cite{nccp,muru}. But this is not unexpected since as we have
mentioned before, the defining conditions of the NC $CP(1)$ models
are different. However, we repeat that {\it{a priori}} there is
nothing inconsistent in our NC model.

Finally, we are ready to discuss the curiosity. It appears that
solutions of the BPS  equations (\ref{bps}) do not satisfy the equation of
motion for $\phi$. A straightforward computation yields the
dynamical equation for $\phi$,
$$
D^{\mu}[(1-\frac{1}{4}\theta^{\alpha\beta}F^{\alpha\beta})D_{\mu}\phi
] +\frac{1}{2}\theta^{\alpha\beta}[iD_{\alpha}\{(D^{\sigma}\phi
)^{\dag}D_{\sigma}\phi D_{\beta}\phi
\}+D_{\beta}\{F_{\alpha\mu}D^{\mu}\phi\}+D^{\mu}\{F_{\alpha\mu}D_{\beta}\phi\}$$
\begin{equation}
-iD_{\alpha}\{(D_{\beta}\phi )^{\dag} D^{\mu}\phi
+(D^{\mu}\phi)^{\dag}D_{\beta}\phi ) D_{\mu}\phi \}
+iD_{\mu}\{(D_{\beta}\phi )^{\dag} D^{\mu}\phi
+(D^{\mu}\phi)^{\dag}D_{\beta}\phi )D_{\alpha}\phi \}]-\lambda
\phi =0. \label{eq}
\end{equation}
(Details of the derivation are provided in the appendix.)
To get $\lambda$, contract by $\phi^{\dag}$ and use $\phi^{\dag}\phi =1$.
For the time being, instead of writing the full equation of motion, we want to
check the consistency of the programme only, that is whether the solution of the BPS equation
satisfies the equation of motion, which they should. Since the BPS equation remains
unchanged here, the  $\theta$-term in the equation of motion should vanish
for those solutions that satisfy  the
BPS equation as well. So we consider the equation of motion in a simplified
setting where the BPS equation is satisfied and only $\theta^{12}$ is non-zero and
obtain for $\lambda$
\begin{equation}
\lambda =\phi^{\dag}D^{i}D_{i}\phi +\frac{1}{2}\theta^{12}\phi^{\dag}D^i(F_{12}D_{i}\phi ).
\label{lam}
\end{equation}
Putting $\lambda$ back in the equation of motion, we get
\begin{equation}
D^{i}D_{i}\phi -\phi^{\dag}D^{i}D_{i}\phi \phi
+\frac{1}{2}\theta^{12}[D^i(F_{12}D_i\phi)
-\phi^{\dag}D^i(F_{12}D_{i}\phi )\phi ]=0. \label{eq2}
\end{equation}
This equation can be rewritten as
\begin{equation}
D^{i}D_{i}\phi -\phi^{\dag}D^{i}D_{i}\phi \phi
+\frac{1}{2}\theta^{12}[\partial^iF_{12}D_i\phi-
\phi^{\dag}\partial^iF_{12}D_{i}\phi \phi +F_{12}(D^{i}D_{i}\phi
-\phi^{\dag}D^{i}D_{i}\phi \phi )]=0. \label{eq3}
\end{equation}
Clearly the last term, that is
$\frac{1}{2}\theta^{12}F_{12}(D^{i}D_{i}\phi
-\phi^{\dag}D^{i}D_{i}\phi \phi )\approx O(\theta^{2})$  and can
be dropped. The term
$\theta^{12}\phi^{\dag}\partial^iF_{12}D_{i}\phi \phi
=\theta^{12}\partial^iF_{12}\phi^{\dag}D_{i}\phi \phi =0$. However
the remaining $O(\theta )$-term,
$\frac{1}{2}\theta^{12}\partial^iF_{12}D_i\phi$  does not vanish.
This is the purported mismatch between the BPS equations and the
full equation of motion. This brings us to the last part - the
collective coordinate analysis, where we can check explicitly the above
conclusions in a simplified setup.

\vskip .5cm
\begin{center}
{\bf{Section VI - Collective coordinate analysis}}
\end{center}
We consider the topological charge $N=1$ sector. As we have
discussed before, expression for the topological current and
subsequently the charge remains same (in our NC $CP(1)$ model) as
that of the ordinary spacetime $CP(1)$ model. This means that we
can use the same parameterizations as before \cite{vij} to
introduce the collective coordinates. As a first approximation,
only the zero mode arising from the global $U(1)$ invariance is
being quantized. In the $O(3)$ nonlinear sigma model, the $N=1$
sector is characterized by \cite{vij}
$$n^a=\{\hat{r}sin(g(r)),cos(g(r))\}~;a=1,2,3$$ with the
constraint $n^an^a=1$ and the boundary conditions $g(0)=0;g(\infty)=\pi$.
Keeping in mind the $O(3)-CP(1)$ duality
and the Hopf map $n^a=\phi^{\dag}\sigma^a\phi$, the soliton profile
in the $CP(1)$ variables is of the form,
\begin{equation}
\phi \equiv
\left (
\begin{array}{c}
 \phi^{1} \\
  \phi^{2}
\end{array}
\right )
=
\left (
\begin{array}{c}
 cos(\frac{g}{2})  \\
  sin(\frac{g}{2})e^{i(\varphi +\alpha (t))}
\end{array}
\right )
\label{col} \end{equation}
where the gauge $(\phi^{1})^*=\phi^{1}$ has been used and $r,\varphi$ refer to the plane polar
 coordinates. $\alpha (t)$ is the collective coordinate. Substituting the above choice (\ref{col}) in the static energy expression in (\ref{en}) leads to,
\begin{equation}
E(r)=(1+\theta \frac{sin(g)g'}{2r})[(g')^2 +\frac{sin^2(g)}{r^{2}}],
\label{cen}
\end{equation}
where $g'=\frac{dg}{dr}$.  In $figure~ (1)$ the effect of the $\theta$-correction is shown where the following  simple form of $g(r)$ is considered,
\begin{equation}
g(r)\approx \pi (1-e^{-\mu r}).
\label{g}
\end{equation}
One can clearly see that with typical values of the parameters, $(\theta =1~,~\mu =1)$
the energy density for the NC case is more sharply peaked. (In reality, $\theta$ should
be smaller.) This assures us of the rationale of our previous Bogolmony analysis. Next
we look in to the equation of motion.

It is straightforward check that the profile (\ref{col}) satisfies the BPS equation as
well as the equation of motion for the ordinary spacetime situation, $\theta =0$. For
$\theta\neq 0$, the BPS equations are once again satisfied since they remain unaltered.
So we concentrate only on the problem term  $\frac{1}{2}\theta^{12}\partial^iF_{12}D_i\phi$
in the equation of motion (\ref{eq3}).  With the particular form of $g(r)$ in (\ref{g}) we
obtain
\begin{equation}
\frac{1}{2}\theta^{12}\partial^iF_{12}D_i\phi =\frac{\theta g'}{4} [\frac{g''}{2r}sin(\frac{g}{2}) -\frac{g'}{2r^{2}}sin(\frac{g}{2}) +\frac{(g')^2}{4r}cos(\frac{g}{2})]
\left (
\begin{array}{c}
 sin(\frac{g}{2}) \\
  -cos(\frac{g}{2})e^{i\varphi }
\end{array}
\right )
\equiv
\left (
\begin{array}{c}
 X(r)\\
 Y(r)
\end{array}
\right ).
\label{col1} \end{equation}
In $Figure~2$ we again use the $g(r)$ given in (\ref{g}) and compare the magnitude of
the above expression with a typical term,
$$\partial^{i}\partial_{i}\phi \equiv
\left (
\begin{array}{c}
 S(r) \\
 T(r)
\end{array}
\right ),
$$
occurring in the equation of motion (\ref{eq3}). For consistency, the expressions
in (\ref{col1}) should have vanished. However, even for $\theta =1$, (which is quite a
large value), the mismatch term is small in an absolute sense.

\vskip .5cm
\begin{center}
{\bf{Section VII - Conclusions}}
\end{center}
In this paper, we have attempted to recover the soliton solutions in the non-commutative
$CP(1)$ model, discovered earlier \cite{nccp}. The $U(1)$ and NC $U(1)$ gauge invariances
in the $CP(1)$ and NC $CP(1)$ models respectively, requires the use of the Seiberg-Witten
map, to convert the NC action  to an action comprising of ordinary spacetime dynamical
variables. The effects of noncommutativity are manifested as interaction terms. For
theoretical as well as technical reasons, we found it convenient to keep the $CP(1)$
constraint unchanged, {\it{i.e.} } without any $\theta$ correction.

From the above action, we construct both the symmetric and canonical forms of the energy
momentum tensor. For only spatial noncommutativity, both the above forms reduce to an
identical (gauge invariant) expression for the static energy.
The Bogomolny analysis yields a lower bound in the energy, hinting at the presence of
a new type of soliton. However, the resulting BPS equations do not match completely with
the full variational equation of motion. The present model is otherwise a perfectly well
defined NC field theory with the expected features. Hence we conclude that {\it{inadequacy in
the definitions of the  energy momentum tensor in an NC field theory is responsible for this
failure.}}
The above phenomena are nicely visualized in a collective coordinate framework.
The above awkward situation clearly demands further study.

Finally, as a future work we mention that inclusion of the Hopf
term, (in the form of Chern-Simons term in $CP(1)$ variables), in
the NC theory would indeed be interesting. The Hopf term was
introduced \cite{vij,cha} to impart anyonic behavior  to the
$CP(1)$ solitons. The exact form of the NC version of the Chern
Simons term is known \cite{gs} - it is a "non-abelian"
generalization of the Chern Simons term. In our formalism,
application of the Seiberg-Witten map will reduce it to the
ordinary Chern Simons term \cite{gs} but there will appear
$O(\theta )$ correction terms since the gauge field of the Chern
Simons term is actually a non-linear combination of the $CP(1)$
variables.

Another interesting problem is the reconstruction of the NC $CP(1)$ model of \cite{nccp}
in our framework.

\vskip .3cm
{\bf {Appendix:}}
To get the $\phi$-equation of motion, we consider variation of $\phi^{\dag}$.
and exploit  the relations,
\begin{equation}
(D^{\mu}\phi )^{\dag}\phi =\phi^{\dag}D^{\mu}\phi =0 ,
\label{rel}
\end{equation}
$$\delta (D^{\mu}\phi )^{\dag}=\delta (\partial^{\mu}\phi^{\dag}+(\phi^{\dag}\partial^{\mu}\phi )\phi^{\dag} )=\partial^{\mu}(\delta\phi^{\dag})+(\delta\phi^{\dag}\partial^{\mu}\phi )\phi^{\dag} +(\phi^{\dag}\partial^{\mu}\phi )\delta\phi^{\dag} $$
\begin{equation}
\delta (D^{\mu}\phi )=\delta (\partial^{\mu}\phi-(\phi^{\dag}\partial^{\mu}\phi )\phi )=
-(\delta\phi^{\dag}\partial^{\mu}\phi )\phi ).
\label{rel1}
\end{equation}
In the action the terms are products of the generic form $\int(D^{\mu}\phi )^{\dag}(D^{\nu}\phi )X(x)$. The variation of $(D^{\mu}\phi )$ will reproduce
$$\int (D^{\mu}\phi )^{\dag}\delta (D^{\nu}\phi )X =-\int (D^{\mu}\phi )^{\dag}\phi (\delta\phi^{\dag}\partial^{\mu}\phi )X =0,$$ by using (\ref{rel}).  Similarly, the variation of $(D^{\mu}\phi )^{\dag}$ will yield
$$\int \delta (D^{\mu}\phi )^{\dag}(D^{\nu}\phi )X =\int (\partial^{\mu}(\delta\phi^{\dag})+(\delta\phi^{\dag}\partial^{\mu}\phi )\phi^{\dag} +(\phi^{\dag}\partial^{\mu}\phi )\delta\phi^{\dag}) D^{\nu}\phi X=-\int\delta\phi^{\dag }D^{\mu}(D^{\nu}\phi X)$$ by partial integration and using (\ref{rel}). the above identities simplifies the computations considerably and leads to the equation (\ref{eq}).
\vskip .5cm
{\bf{Acknowledgements:}} It is a pleasure to thank Professor Hyun Seok Yang for fruitful
correspondence. Also I thank Professor Avinash Khare for a helpful discussion.
Lastly I am indebted to Bishwajit Chakraborty for a free access to his notes on $CP(1)$ model.

\end{document}